# Magnetotransport of CeRhIn$_5$

A.D. Christianson, A.H. Lacerda
National High Magnetic Field Laboratory, Los Alamos National Laboratory,
Los Alamos, NM 87545, USA

M.F. Hundley, P.G. Pagliuso, and J.L. Sarrao
Condensed Matter and Thermal Physics Group, Los Alamos National Laboratory,
Los Alamos, NM 87545, USA

## Abstract

We report measurements of the temperature-dependent anisotropic resistivity and in-plane magnetoresistance on single crystals of the tetragonal heavy-fermion antiferromagnet ($T_N$ = 3.8 K) CeRhIn$_5$. The measurements are reported in the temperature range 1.4 K to 300 K and in magnetic fields to 18 tesla. The resistivity is moderately anisotropic, with a room-temperature c-axis to in-plane resistivity ratio $\rho_c/\rho_a$(300 K) = 1.7. $\rho$(T) measurements on the non-magnetic analog LaRhIn$_5$ indicate that the anisotropy in the CeRhIn$_5$ resistivity stems predominately from anisotropy in Kondo-derived magnetic scattering. In the magnetically ordered regime an applied field H reduces $T_N$ only slightly due to the small ordered moment (0.37$\mu_B$) and magnetic anisotropy. The magnetoresistance (MR) below $T_N$ is positive and varies linearly with H. In the paramagnetic state a positive MR is present below 7.5 K, while a high-field negative contribution is evident at higher temperatures. The positive contribution decreases in magnitude with increasing temperature. Above 40 K the positive contribution is no longer observable, and the MR is negative. The low-T positive MR results from interactions with the Kondo-coherent state, while the high-T negative MR stems from single-impurity effects. The H and T-dependent magnetotransport reflects the magnetic anisotropy and Kondo interactions at play in CeRhIn$_5$.

# I.   Introduction

Transport measurements in large applied magnetic fields provide an exceptionally useful means of probing the electronic and thermodynamic properties of heavy fermion compounds. This stems from the magnetic origin of the interactions responsible for the mass-enhanced ground state.[1] As such, the resistivity of a heavy fermion system is altered by an applied magnetic field in fundamentally different ways when the compound is in a magnetically ordered, Kondo-coherent or single-impurity regime. Field-dependent measurements can also provide information regarding the importance of magnetic fluctuations and the proximity to low-temperature magnetic instabilities in the coherent regime.  Although no complete microscopic theory is available to fully model the transport and thermodynamic properties of a heavy fermion system, a number of theoretical treatments are available that qualitatively describe the key features of a system's field-dependent behavior.[2,3] From the experimental point of view, a wide range of phenomena can be observed when applying a magnetic field at low temperature.[4] Among archetype heavy fermion compounds, $UPt_3$ and $CeRu_2Si_2$ exhibit highly anisotropic magnetotransport behavior and remarkable field-induced metamagnetic transitions at 20 T,[5,6] and 8 T,[7] respectively, while $UBe_{13}$, exhibits negative magnetoresistance above its superconducting transition temperature.[8] Ultimately, a heavy fermion system's field-dependent properties are determined by RKKY and Kondo interactions,[9] with the relative importance of these two interactions influenced by magnetic and structural anisotropies.

A new family of Ce-based heavy fermions was recently discovered that exhibits a complex phase diagram that challenges our understanding of correlated electron physics.[10,11] This family has a generalized chemical formula, $Ce_mM_nIn_{3m+2n}$, where $M$ is Rh, Ir, or Co. All compounds investigated to-date (m = 1, 2 and n = 1), except cubic $CeIn_3$, crystallize in a tetragonal structure (space group P4/mmm).[12] The most notable properties in this series of compounds include ambient-pressure magnetic order ($T_N$ = 3.8 K) and pressure-induced superconductivity ($T_c$ = 2.1 K at 16 kbar pressure) in CeRhIn$_5$,[10,13,14] and unconventional[15] ambient-pressure superconductivity in both CeIrIn$_5$ ($T_c$ = 0.4 K)[16] and CeCoIn$_5$ ($T_c$ = 2.3 K);[17] the transition temperature for CeCoIn$_5$ is the highest ambient-pressure $T_c$ reported to date for a heavy fermion superconductor. This new family of compounds offers the opportunity to explore the importance of tuned dimensionality on magnetic, Kondo, and superconducting groundstates.

CeRhIn$_5$ has attracted considerable attention due to its unusual pressure-temperature phase diagram.[10] Ce heavy fermion systems that order antiferromagnetically typically exhibit a P-T phase diagram wherein applied pressure acts to smoothly reduce the Neel temperature $T_N$ to zero at a critical pressure $P_c$, with superconductivity occurring over a range of pressure centered at $P_c$. The P-T phase diagram of the cubic member of the Ce$_m$M$_n$In$_{3m+2n}$ family (CeIn$_3$) displays this behavior, with an ambient-pressure ordering temperature $T_N$ = 10 K, a slightly enhanced Sommerfeld coefficient of 100 mJ/mole-K$^2$, and a critical pressure $P_C$ = 23 kbar.[18,19] The CeRhIn$_5$ P-T phase diagram is quite different. At ambient pressure, CeRhIn$_5$ orders antiferromagnetically at 3.8 K. Applied hydrostatic pressure acts to very slightly increase $T_N$ until magnetic order becomes unobservable near 16 kbar, at which point superconductivity appears at 2.1 K.[10] Specific heat measurements indicate an enhanced Sommerfeld coefficient of roughly 420 mJ/mole-K$^2$ below 10 K;[10] for a single impurity system,[20] this corresponds to a Kondo temperature of roughly 10 K. CeRhIn$_5$ has a quasi-2D structure that is composed of alternating layers of the cubic heavy fermion antiferromagnet CeIn$_3$ and a transition-metal layer composed of RhIn$_2$. As such, dimensionality may play a role in the interactions that produces the unusual P-T phase diagram exhibited by CeRhIn$_5$. This is born out by nuclear quadrupolar resonance[21] and neutron scatting measurements[22] which indicate that the magnetic moments lie in the basal plane of the tetragonal structure with a spiral along the c-axis, with a reduced magnetic moment of 0.37 Bohr magnetons ($\mu_B$). However, recent inelastic neutron scattering experiments indicate some degree of 3-dimensional behavior for $CeRhIn_5$.[23] Clearly, further measurements are needed to fully elucidate the influence of dimensionality on the physical properties of $CeRhIn_5$.

In order to enhance our understanding of the ground state properties of $CeRhIn_5$, we have measured the anisotropic resistivity of this compound as a function of magnetic field and temperature. The resistivity is moderately anisotropic, with a room-temperature c-axis to in-plane resistivity ratio $\rho_c/\rho_a$ of 1.7. This ratio changes markedly with decreasing temperature, and at 4 K the in-plane resistivity is larger than the out-of-plane resistivity by 80%. The antiferromagnetic transition at 3.8 K produces an inflection point in $\rho$. With application of magnetic field, the transition moves to slightly lower temperatures, with a field-dependence that depends upon the direction of the applied field. The magnetoresistance (MR) also depends significantly upon the direction of the applied field. The MR is positive in the magnetically ordered state and varies linearly with applied field. At moderate temperatures, we observe positive contribution to the MR that is characteristic of a Kondo system in the coherent regime. At higher temperature, this positive MR gives way to a negative contribution characteristic of a single impurity Kondo system.

## II.    Experimental details

Single crystals of $CeRhIn_5$ were grown from an In flux method[24] as described previously.[12] The deleterious influence of residual In flux on low-T transport measurements (the superconducting transition for In occurs at 3.4 K) necessitates careful sample surface polishing to remove any possible In contamination. The polished single crystal samples were orientated by using Laue x-ray diffraction to determine the crystallographic in-plane (a-axis) and out-of-plane (c-axis) directions. Finally, the resistance of each sample that was slated for use in MR measurements was measured down to 2 K to ensure that no extrinsic superconductivity contamination was evident at 3.4 K due to surface In. The samples that passed this screening process had residual resistivity ratios [RRR = $\rho$(300 K)/$\rho$(4 K) $\approx$ 100] that were similar to those reported previously.[25]

All resistivity measurements reported here were made with a conventional four-probe sample configuration in which silver conductive paint or epoxy was used to make sample contacts. Sample resistances were measured with a low-frequency ac bridge. The in-plane and out-of-plane resistivities were determined on oriented samples via the Montgomery and anisotropic van der Pauw methods.[26] The transverse magnetoresistance was measured with current applied along an a-axis, and the applied field oriented perpendicular to the measurement current (i.e., either in the other a-axis or along the c axis). The transport measurements were carried out in a variable flow cryostat capable of temperatures from 1.4 K to 325 K. To avoid magnetoresistance effects in the Cernox thermometer used to determine and control sample temperature, temperatures below 3 K were stabilized by controlling the [4]He vapor pressure.

## III.  Results

The temperature-dependent resistivities of CeRhIn$_5$ and LaRhIn$_5$ in, and perpendicular to, the basal plane, are shown in Fig. 1a. The data for CeRhIn$_5$ indicate that this compound is moderately anisotropic; $\rho_c$ is roughly 70% larger than $\rho_a$ at room temperature. Below 325 K the resistivity falls with decreasing temperature in both directions, and both $\rho_a$ and $\rho_c$ exhibit shoulder-like features between 50 and 100 K. Both resistivities fall-off more rapidly at lower temperatures. $\rho_a$ and $\rho_c$ cross at 30 K, and the a-axis resistivity is larger than the c-axis resistivity down to 1 K. In comparison, the resistivity of $LaRhIn_5$ (the non-magnetic analog of $CeRhIn_5$) varies linearly with temperature below 300 K, and saturates to a value near 1 $\mu\Omega$ cm below 20 K. The $LaRhIn_5$ c-axis resistivity is greater than the in-plane resistivity at all temperatures, and the anisotropy ratio $\rho_c/\rho_a$ is nearly T-independent. The 300 K anisotropy ratio $\rho_c/\rho_a$ = 1.2 for $LaRhIn_5$ suggests that the non-magnetic electronic anisotropy inherent to the $RMIn_5$ structure is relatively small.

The temperature-dependent magnetic scattering component ($\rho_{mag} = \rho_{Ce} - \rho_{La}$) of the $CeRhIn_5$ in-plane and c-axis resistivities are presented in Fig. 1b. After removing the electron-phonon scattering contribution to $\rho_{Ce}$, the magnetic resistivity in both crystallographic directions varies as $\rho \propto -\ln(T)$ at high temperatures and drops sharply below 50 K; this T-dependence is characteristic of Kondo lattice compounds.[1] The resistivity in the vicinity of the 3.8 K AFM transition is shown in the inset to Fig. 1b. A clear change in magnetic scattering is evident in both $\rho_a$ and $\rho_c$ near $T_N$. The transport anisotropy ratio $\rho_a^{mag} / \rho_c^{mag}$ is plotted as a function of temperature in Fig. 2. Near room temperature the magnetic resistivity is moderately anisotropic ($\rho_a^{mag} / \rho_c^{mag}$ at 300 K is 0.6), and the ratio exhibits a gradual evolution from a high-T regime where $\rho_a^{mag} / \rho_c^{mag} < 1$ to a low-T regime where $\rho_a^{mag} / \rho_c^{mag} > 1$. The magnetic resistivities cross at 30 K. $\rho_c^{mag}$ is smaller than $\rho_a^{mag}$ down to the lowest measurement temperature (1.4 K), and there is no evidence for any change in $\rho_a^{mag} / \rho_c^{mag}$ at or below $T_N$.

We now turn to an examination of the influence of applied magnetic fields on the T-dependent in-plane resistivity. The resistivity as a function of temperature in a field of 18 tesla is displayed in Fig. 3, and compared to the zero-field $\rho_a$ data. In Fig. 3a the magnetic field is applied parallel to the basal plane and perpendicular to the current. A positive MR is evident at low temperatures, with the magnitude of the effect diminishing with increasing temperature. Above roughly 50 K, no difference is discernable between $\rho(H = 0)$ and $\rho(18 \text{ T})$. The inset to Fig. 3a shows $\rho(T)$ in fields of 0, 10, 15, and 18 T in the vicinity of $T_N$; in this temperature regime the applied fields appear to uniformly increase the resistivity below 4 K. The H-dependent AFM ordering temperature can be determined by finding the location of the inflection point in $\rho$ that marks $T_N$. The arrows in the inset denote $T_N(H)$. The transition moves downward monotonically with temperature; in 18 T the inflection point occurs at 3.35 K, corresponding to the rate $dT_N/dH_\parallel = -25$ mk/T. The a-axis resistivity for a field applied parallel to the c-axis is shown in Fig. 3b. In this field orientation, the low-T magnetoresistance is also positive, but the 18 T MR crosses zero at 16 K, and becomes large and negative at higher temperatures. This negative MR effect reaches a maximum value nearly 30 K. At higher temperatures the negative MR diminishes in magnitude, approaching zero at 100 K. The inset to Fig. 3b depicts $\rho(T)$ in fields of 0, 5, 10, 15, and 18 T in the vicinity of $T_N$; as with the in-plane field orientation, the applied field uniformly increases $\rho_a$ below 5 K. The field also decreases the AFM transition temperature, but at a faster rate than for fields oriented in the basal plan. In 18 T the applied field drops $T_N$ to 3.0 K; this corresponding to a rate $dT_N/dH_\perp = -35$ mk/T, a value that is in good agreement with independent rate measurements from heat-capacity.[27,32]

The field-dependent in-plane magnetoresistance $\Delta\rho_a(H) = \rho_a(H) - \rho_a(H=0)$ at constant temperature is depicted in Figures 4 ($H \parallel a$) and 5 ($H \parallel c$) for CeRhIn$_5$. With the field applied in the basal plane the MR below 10 K (Fig. 4a) exhibits two distinct regimes. At 1.4 K the $\Delta\rho_a(H)$ varies linearly with H throughout the measured field range ($H \leq 18$ T), while for $T > T_N$ the MR grows in magnitude and exhibits some curvature below 5 T. At 7.5 K the MR varies as $H^{1/2}$ above 1 tesla, and it saturates above 15 T. At still higher temperatures (Fig. 4b) $\Delta\rho_a(H)$ displays a maximum at a field $H_{max}$ that occurs near 12 tesla. The rise in the MR at low fields is suppressed as the temperature is increased, and the overall magnitude of the magnetoresistance diminishes as well. The relative magnetoresistance in 18 tesla, defined as $[\rho_a(H) - \rho_a(H=0)]/\rho_a(H=0)$, is plotted as a function of temperature in the inset to Fig. 4a. The relative MR is nearly zero above 20 K, and grows markedly below 10 K in large measure due to the sharp drop in $\rho_a(T, H = 0)$ that stems from the onset of coherence. In contrast, the magnetoresistance of the non-magnetic analog LaRhIn$_5$ displaying a standard metal-like positive MR that varies as $H^2$ and diminishes in magnitude with increasing temperature.

The in-plane magnetoresistance of CeRhIn$_5$ with H applied along the c axis is depicted in Fig. 5. For $T \leq 7.5$ K (Fig. 5a) the results are qualitatively similar to those for H ∥ a. Below $T_N$ $\Delta\rho_a(H)$ grows linearly with H, and a small change in slope is evident near 2.5 T. Above $T_N$ the high-field MR grows as $H^\alpha$ with $\alpha < 1$. At 7.5 K $\Delta\rho_a(H)$ varies as $H^{1/2}$ throughout the measured field range and it is approaching saturation at 18 T. For $T \geq 7.5$ K (Fig. 5b) the MR is quite different from the low-T behavior. The $H^{1/2}$ behavior present at 7.5 K evolves into a peak in $\Delta\rho_a(H)$ at 10 K that occurs between 5 and 10 T, and the MR decreases markedly at still higher fields. Above 20 K, the low-H positive MR is no longer in evidence and the negative MR contribution predominates. The MR is negative above 30 K at all fields, and the overall magnitude of the negative MR decreases with increasing temperature.

Taken as a whole, the temperature and magnetic-field dependent $\rho_a$ data presented in Figs. 3-5 suggest that there are three field-dependent transport regimes in CeRhIn$_5$. The first, in the magnetically ordered state, exhibits a nearly linear-in-H MR that does not show any sign of field saturation at 18 T (we note that at least 40 T is required to field-polarize the AFM state).[28] The second regime resides in the paramagnetic state just above T$_N$. In this regime the MR is positive and exhibits a tendency to saturate near 20 tesla. The third regime occurs at temperatures above 10 K and at high fields where a negative MR contribution comes into play that initially produces a maximum in $\Delta\rho_a(H)$. At still higher temperatures the positive MR disappears and the negative contribution dominates the field-dependent transport. Magnetic anisotropy influences the detailed nature of the field-dependent transport. The influence of the high-T negative MR contribution is largest with the field applied perpendicular to the basal plane. As such, the peak field H$_{max}$ is largest with the field applied in the basal plane, and the MR is more negative for H ∥ c.

## IV.    Discussion

The anisotropy in the zero-field resistivity data, and the complex magnetic-field and temperature dependence of the a-axis magnetoresistance are the most prominent features of these CeRhIn$_5$ magnetotransport data. How do these features reflect the tetragonal crystal structure, the Kondo and crystal-field interactions, and the RKKY-mediate antiferromagnetic order? Before answering these questions, we first must examine the influence that lattice anisotropy has on the electronic and magnetic structure in the CeRhIn$_5$.

The CeRhIn$_5$ unit cell is composed of cubic CeIn$_3$ building-blocks that are separated by RhIn$_2$ layers. Full-potential band structure calculations[29] indicate that the electronic structure of CeRhIn$_5$ and LaRhIn$_5$ reflects the quasi-2D nature of the tetragonal unit cell. The band structure exhibits a number of bands that cross the fermi energy E$_f$, producing three fermi surfaces. Only the first, containing hole-like orbits, is relatively isotropic. Reflecting CeRhIn$_5$'s planar structure, the second and third surfaces are composed of corrugated cylindrical electron- and hole-like orbits that extend along the c axis. de Haas-van Alphen (dHvA) measurements detect extremal orbits that are consistent with the band-structure calculations.[25,29] In addition, the Hall effect in both CeRhIn$_5$ and LaRhIn$_5$ is anisotropic and strongly temperature dependent,[30] providing clear evidence for competing electron and hole carriers. The fact that the Hall effect in CeRhIn$_5$ and LaRhIn$_5$ are quite similar indicates that they share the same anisotropic electronic structure, and that the f-electrons in CeRhIn$_5$ are localized.[31] Hence, based both on measurements and calculations, the layered structure of CeRhIn$_5$ is reflected in the compound's complex electronic structure.

The magnetic structure of the antiferromagnetic ground state also reflects CeRhIn$_5$'s layered nature. The magnetic moments that order at T$_N$ = 10 K in CeIn$_3$ are commensurate with the cubic lattice.[19] In contrast, the magnetic moments in CeRhIn$_5$ are found to lie completely within the basal plane, and they form an incommensurate spiral along the c-axis.[21,22] Field-dependent specific-heat[32] and dHvA[25] measurements indicate that fields oriented along the c-axis gradually reduce the ordering temperature without altering this incommensurate structure. Fields applied within the basal plane strongly influence the magnetic structure, producing a complex H-T phase diagram.[28,32,33] Below 3 K a field of 2 tesla transforms the magnetic structure to one that is commensurate with the lattice, while a third state is also present near 3.5 K. The onset ordering temperature is much less field-dependent than for H ∥ c. In the paramagnetic regime the magnetic susceptibility χ exhibits a factor-of-two anisotropy between χ$_a$ and χ$_c$.[10] This anisotropy stems from the splitting of the J = 5/2 manifold under the influence of tetragonal crystalline electric fields. The crystal field level scheme that describes $\chi_a(T)$ and $\chi_c(T)$ in the paramagnetic state[34] includes a $\Gamma_7$ doublet groundstate (composed predominately of the $|\pm 5/2\rangle$ spin state), a first-excited $\Gamma_7$ doublet (predominately $|\pm 3/2\rangle$) at 6 meV, and the last state, a spin-½ $\Gamma_6$ doublet, located 13 meV above the $\Gamma_7$ groundstate. This level scheme is also consistent with preliminary inelastic neutron scattering results that find a broad crystal-field feature at an energy of 6 to 7 meV corresponding to transitions amongst these three levels.[35] Armed with this information concerning $CeRhIn_5$'s electronic and magnetic structures, as well as the crystal-level scheme, we can now examine the underlying mechanisms responsible for the magnetotransport features exhibited by $CeRhIn_5$.

The modest transport anisotropy exhibited by $LaRhIn_5$ ($\rho_c/\rho_a \approx 1.2$) indicates that the quasi-2D electronic structure does not translate into transport anisotropy. Conventional electron-phonon scattering also appears to be weakly influenced by the planar 115 structure as well. The absence of significant anisotropy in the resistivity of $LaRhIn_5$ indicates that the anisotropy in the $CeRhIn_5$ resistivity stems from magnetic scattering. Both the a-axis and c-axis magnetic resistivities of $CeRhIn_5$ display temperature-dependencies that are characteristic of a Kondo-lattice compound. The complex T-dependent anisotropy between the a and c axis magnetic resistivities is reminiscent of that seen in many other heavy-electron systems. For example, The a-axis and c-axis resistivities in the tetragonal compounds $CeRu_2Si_2$ and $CeNi_2Ge_2$ also cross in a manner reminiscent of $CeRhIn_5$.[36,37] There are also a number of other f-electron compounds that exhibit an anisotropic $\rho_{mag}$ but without any crossing of the $\rho_a$ and $\rho_c$ resistivities. Systems that fall into this second class include the hexagonal compound $UPt_3$,[38] orthorhombic $CeCu_6$,[39] and the tetragonal compounds $CePt_2Si_2$,[40] $CePd_2Si_2$,[41] and $CeCu_2Si_2$.[42] As with $CeRhIn_5$, $\rho_a$ and $\rho_c$ never differ by more than a factor of 2 in these systems. These resistivity anisotropies can be explained by considering the nature of the scattering relaxation rates that are produced when resonant Kondo scattering is influenced by anisotropic crystal-field levels.[43,44] This modeling describes successfully the anisotropy evidenced by a wide variety of Ce compounds.[40,43,44,45] As such, it seems reasonable to conclude that the magnetic resistivity anisotropy in CeRhIn$_5$ is a reflection of anisotropic carrier scattering due to the influence of the crystal fields.

The influence of an applied magnetic field on the resistivity in the vicinity of the antiferromagnetic transition is depicted in the insets to Figs. 3a and 3b. The zero-field AFM order at T$_N$ = 3.8 K gives rise to an inflection point in the resistivity, indicating that magnetic order alters the transport in at most a modest way. The absence of any abrupt change in the ratio $\rho_a/\rho_c$ at T$_N$ indicates that the onset of magnetic order influences spin-wave scattering isotropically; this is consistent with inelastic neutron scattering measurements[23] which indicate that there is no 3-D to 2-D crossover prior the onset of long-range order and that the magnetic system is predominately three-dimensional. Specific-heat[32] measurements show that a magnetic field applied in the basal plane will split the antiferromagnetic transitions into three separate transitions. Preliminary neutron-diffraction measurements[33] indicate that these transitions are associated with an evolution in the zero-field magnetic structure. No such splitting of the antiferromagnetic transition signature is evident in the resistivity data shown in the Fig. 3a inset. This may be due to the relatively small change in carrier scattering that will occur when the system is transformed from one magnetic structure to the next, and as such the resistivity inflection points for the separate transitions may be indistinguishable. However, the inflection point, as indicated by the arrows in the inset, decreases very gradually (dT$_N$/dH$_\parallel$ = -25 mk/T) with magnetic field. Specific-heat measurements[27,32] with the magnetic field applied along the c axis indicate that while the field does not alter the magnetic structure it does have a stronger influence on $T_N$; this is consistent with the more rapid field-induced decrease in the inflection-point temperature ($dT_N/dH_\perp$ = -35 mk/T) evident in the data displayed in the Fig. 3b inset.

We now turn to the a-axis magnetoresistance in the paramagnetic state. The data reflect two field/temperature regimes. At low H and T the MR is positive, while at high H and T the MR exhibits a negative contribution. A low-temperature positive MR is a common feature of Kondo systems that are in, or are approaching, a coherent fermi-liquid state.[3,46] A positive magnetoresistance has been reported in both $CeAl_3$ [47,48] and $CeRu_2Si_2$ [49] at low temperatures, and in $UBe_{13}$ under pressure.[50] The positive MR appears to stem from the magnetic field's interaction with the fermi liquid state.[3] Field-dependent transport coefficient calculations for a Kondo-lattice model[3] indicate that the applied field will act to smear out and shift the Abrikosov-Suhl resonance present in the electronic density of states. In essence, the applied field suppresses the Kondo effect by polarizing the local multiplet. For Ce compounds the Kondo resonance occurs just above the fermi energy $E_F$. The model calculation indicate that an applied field shifts the resonance towards $E_F$, and in so doing, increases $N(E_F)$. At a characteristic field $H^* = k_B T_K / \mu_B$ the resonance is centered at $E_F$; for still larger fields the resonance structure is smeared out, and progressively disappears. The field-induced increase in $N(E_F)$ for $H < H^*$ leads to enhanced s-f scattering and a positive magnetoresistance; the MR will reach a maximum at $H^*$. Given a 10 K Kondo temperature for $CeRhIn_5$ (estimated from the 420 mJ/mole-$K^2$ Sommerfeld coefficient),[10] the predicted characteristic field is roughly 15 tesla. This estimate is quite close to the characteristic fields observed for fields aligned within the basal plane ($H_a^* \approx 12$ T) and along the c axis ($H_c^* \approx 8$ T). Hence, the positive low-T MR effect appears to be a manifestation of field-induced destruction of Kondo coherence. In contrast, the negative MR occurs in the single-impurity regime, where the zero-field resistivity varies logarithmically with temperature ($d\rho/dT < 0$). In the single-impurity regime an applied field reduces incoherent Kondo scattering, producing a negative MR.[51,52] In this regime the MR is known to scale with the induced magnetization M as $\Delta\rho/\rho_o \propto -M^2$.[52] Hence, a plot of $\Delta\rho/\rho_o(M)$ for all H and T should fall onto a single, universal curve. A careful analysis of the magnetoresistance data for all temperature that exhibit a hint of a negative MR is made problematic by the interaction between the low-H positive effect and the high-H negative contribution. Nonetheless, this single-impurity analysis is possible with H ∥ c for T ≥ 40 K, as the MR shows no positive contribution in this temperature range. These data are plotted as a function of $M^2$ in Fig. 7. The data scale as expected, falling on a common line and varying as $M^2$. Hence, the negative high-temperature MR contribution appears to be a simple-impurity effect. At these temperature the applied field reduces incoherent Kondo scattering, giving rise to a negative MR. The detailed nature of the MR, and in particular the anisotropy evident in H* and the weaker negative contribution for H ∥ a are an indication that the magnetic anisotropy evident in $\rho_{mag}$ also influences the detailed balance between coherent and incoherent MR effects in CeRhIn$_5$. As such, the temperature-dependent CeRhIn$_5$ magnetotransport reflects the prevalent Kondo regime (coherent at low T, single-impurity at high-T) as well as the magnetic anisotropy stemming from the nature of the crystal-field levels.

## V.  Conclusions

Anisotropy plays a critical role in determining the physical properties of CeRhIn$_5$. Tetragonal crystalline electric fields split the J = 5/2 manifold into three doublets whose anisotropy influences both the magnetic susceptibility[10] and the zero-field resistivity. The RhIn$_2$ spacer layer alters the c-axis magnetic exchange sufficiently to produce antiferromagnetism with an incommensurate spiral spin structure.[21,22] Dimensionality effects are also evident in the way an applied H-field alters this spin arrangement.[32,33] Both dHvA[25,29] and Hall effect[30] measurements indicate that $CeRhIn_5$'s electronic structure is strongly two-dimensional. And, finally, while the overall field and temperature-dependent magnetoresistance in the paramagnetic regime is predominately determined by Kondo-lattice and single-impurity Kondo interactions, the detailed interplay between positive and negative MR contributions also manifests the impact of anisotropy on the magnetotransport. Together, all of these features bolster the claim that the unique pressure-temperature $CeRhIn_5$ phase-diagram[10] stems, in part, from dimensionality effects. Magnetotransport measurements are underway on the ambient-pressure superconducting members of the 115 family ($CeIrIn_5$ and $CeCoIn_5$) to determine the role of anisotropy in these systems as well.

**Acknowledgements**

We gladly acknowledge useful discussions with J.M. Lawrence (University of California, Irvine) and S. Kern (Colorado State University). The work at Los Alamos National Laboratory was preformed under the auspices of the U.S. Department of Energy. The work at the National High Magnetic Field Laboratory, Los Alamos Facility was performed under the auspices of the National Science Foundation, the State of Florida, and the U.S. Department of Energy. One of us (A.D.C.) would like to acknowledge partial support from the Manuel Lujan Jr. Neutron Scattering Center (Los Alamos National Laboratory).

# Figure Captions

Fig. 1. (a) In-plane (solid lines) and c-axis (dashed lines) temperature-dependent resistivities of CeRhIn$_5$ and the non-magnetic analog LaRhIn$_5$. (b) The in-plane (solid line) and c-axis (dashed line) magnetic resistivity ($\rho_{mag} = \rho_{Ce} - \rho_{La}$) of CeRhIn$_5$. The data near T$_N$ are highlighted in the inset, with the arrow positioned at T$_N$.

Fig. 2. Temperature-dependent magnetic resistivity anisotropy ratio ($\rho_a^{mag}/\rho_c^{mag}$) of CeRhIn$_5$.

Fig. 3. In-plane temperature-dependent resistivity in an applied magnetic field. In (a) the magnetic field is applied in the basal plane (perpendicular to the current). The inset displays an expanded view near the AF transition (the curves correspond to $\mu_o$H = 0, 10, 15, and 18 T). The arrows mark the inflection point in $\rho$ (located at 3.7, 3.45, 3.39, and 3.35 K). In (b) the field is applied along the c-axis. The curves in the inset correspond to $\mu_o$H = 0, 5, 10, 15, and 18 T. The inflection points are located at 3.8, 3.7, 3.6, 3.3, and 3.1 K.

Fig. 4. In-plane field-induced change in resistivity $\Delta\rho = (\rho(H) - \rho(0))$ (H $\parallel$ a). The low-T behavior is featured in (a). The inset shows the magnetoresistance $\Delta\rho/\rho(0)$ at 18 T for T < 30 K. The value for the 1.4 K magnetoresistance at 18 T ($\Delta\rho/\rho(0)$ = 8.7) for is not displayed due to its large magnitude. The high temperature behavior of $\Delta\rho$ is displayed in (b).

Fig. 5. In-plane field-induced change in resistivity $\Delta\rho = (\rho(H) - \rho(0))$ (H $\parallel$ c). The low-T behavior is featured in (a). The high temperature behavior of $\Delta\rho$ is displayed in (b).

Fig. 6. In-plane magnetoresistance $\Delta\rho^a/\rho_o$ at 40 and 80 K plotted as a function of the magnetization squared. H is applied along c, and the M$^2$ units are Bohr magnetons per Ce atom.

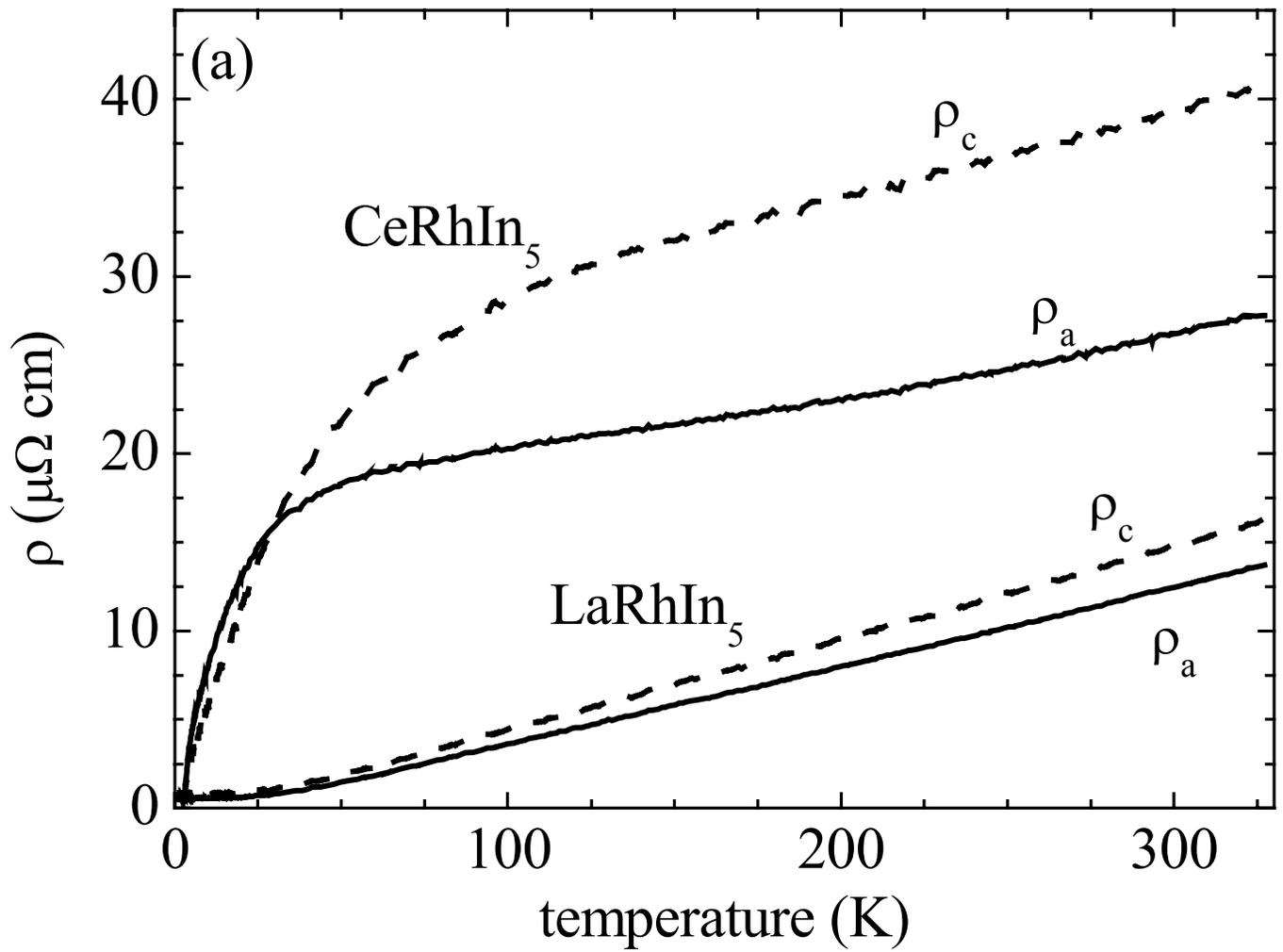

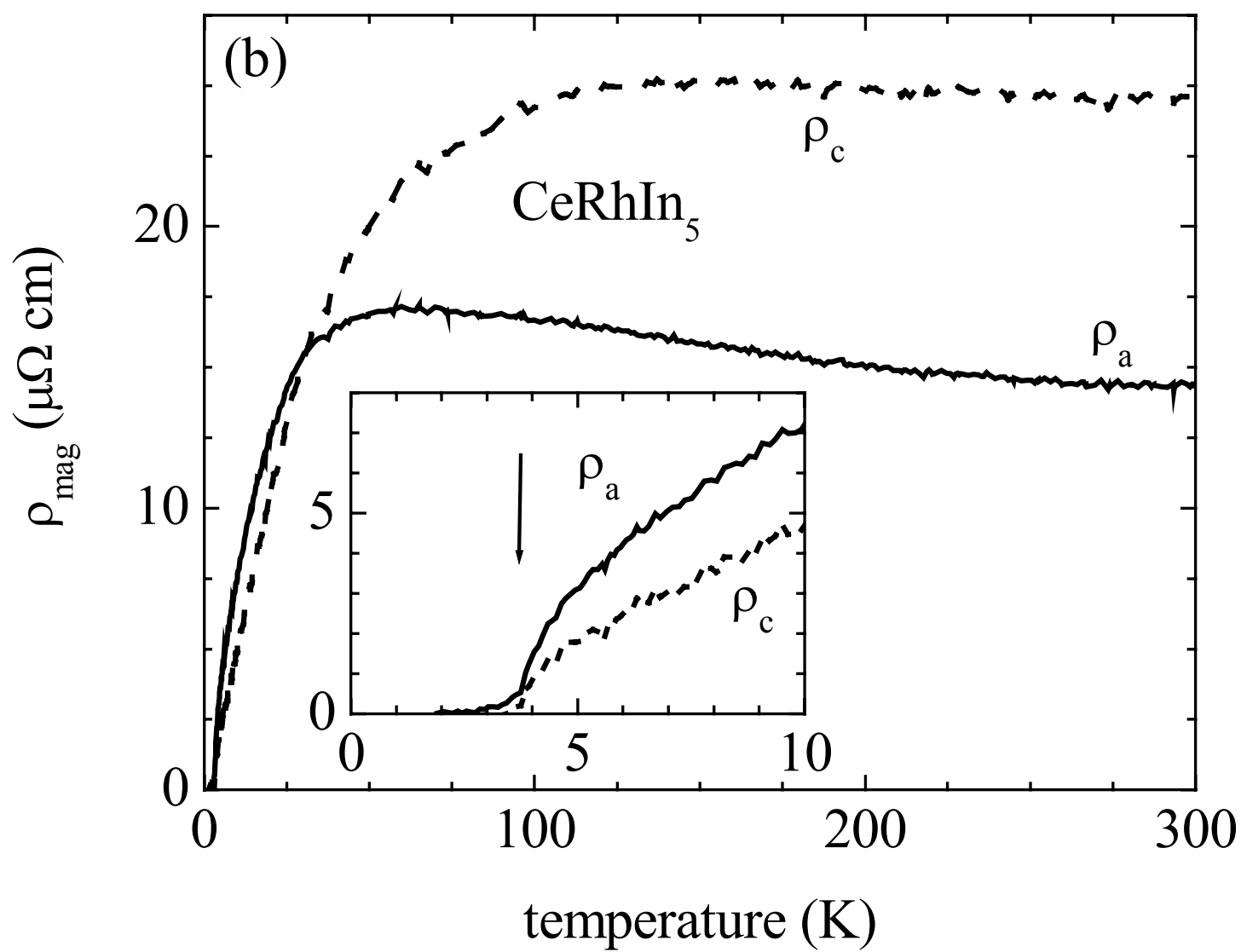

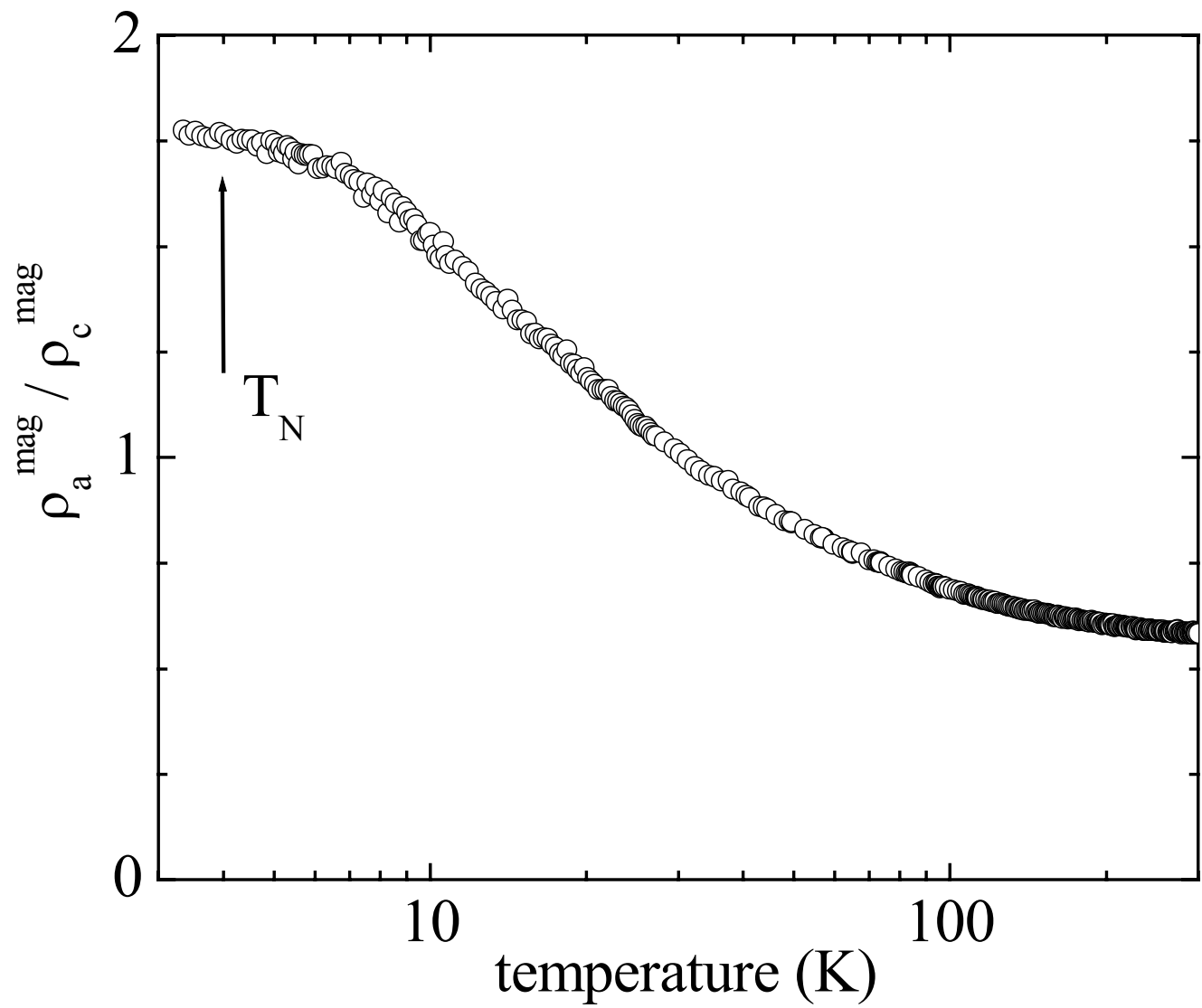

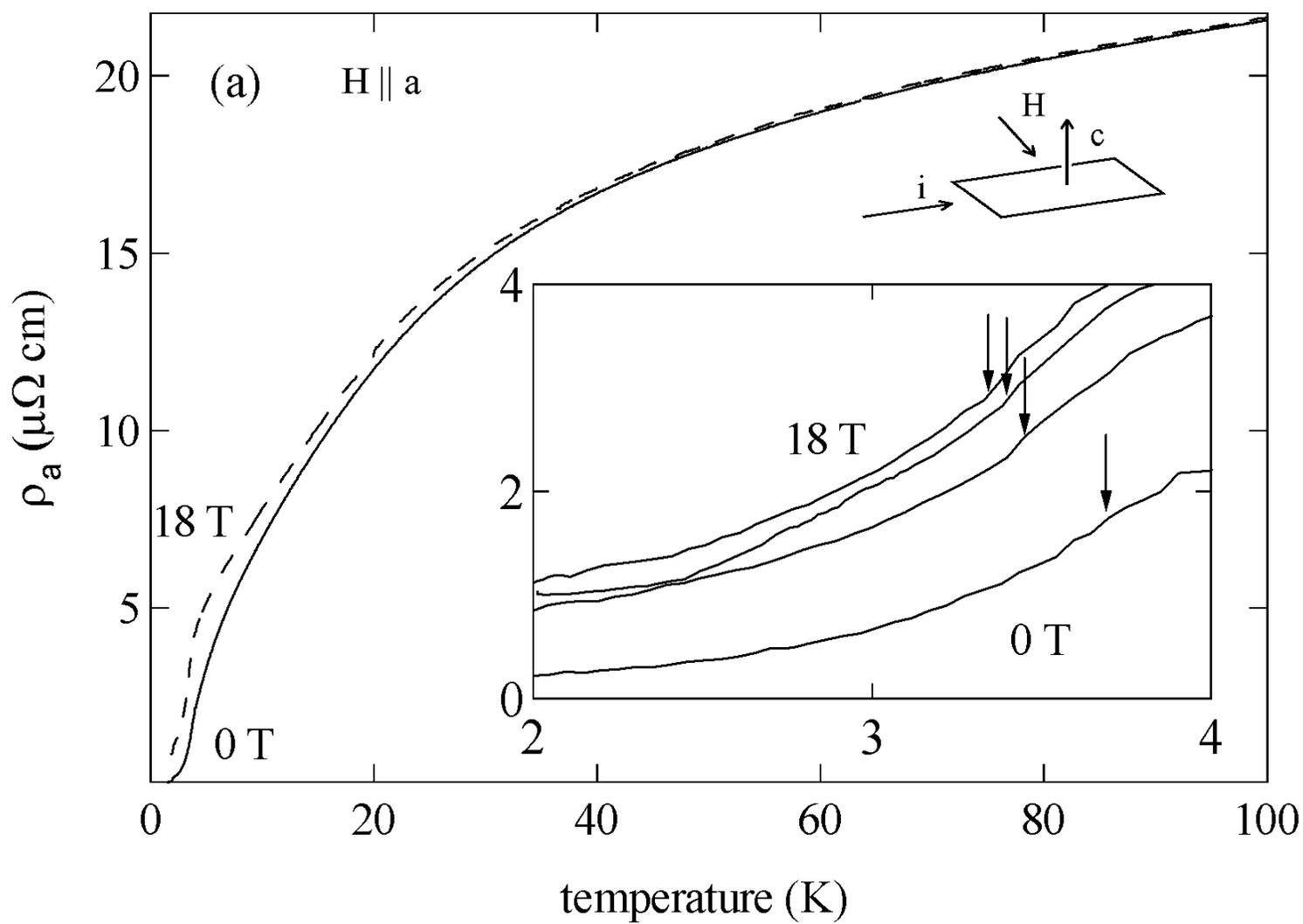

(a)   H ∥ a

18 T

0 T

18 T

0 T

$\rho_a$ ($\mu\Omega$ cm)

temperature (K)

H

c

i

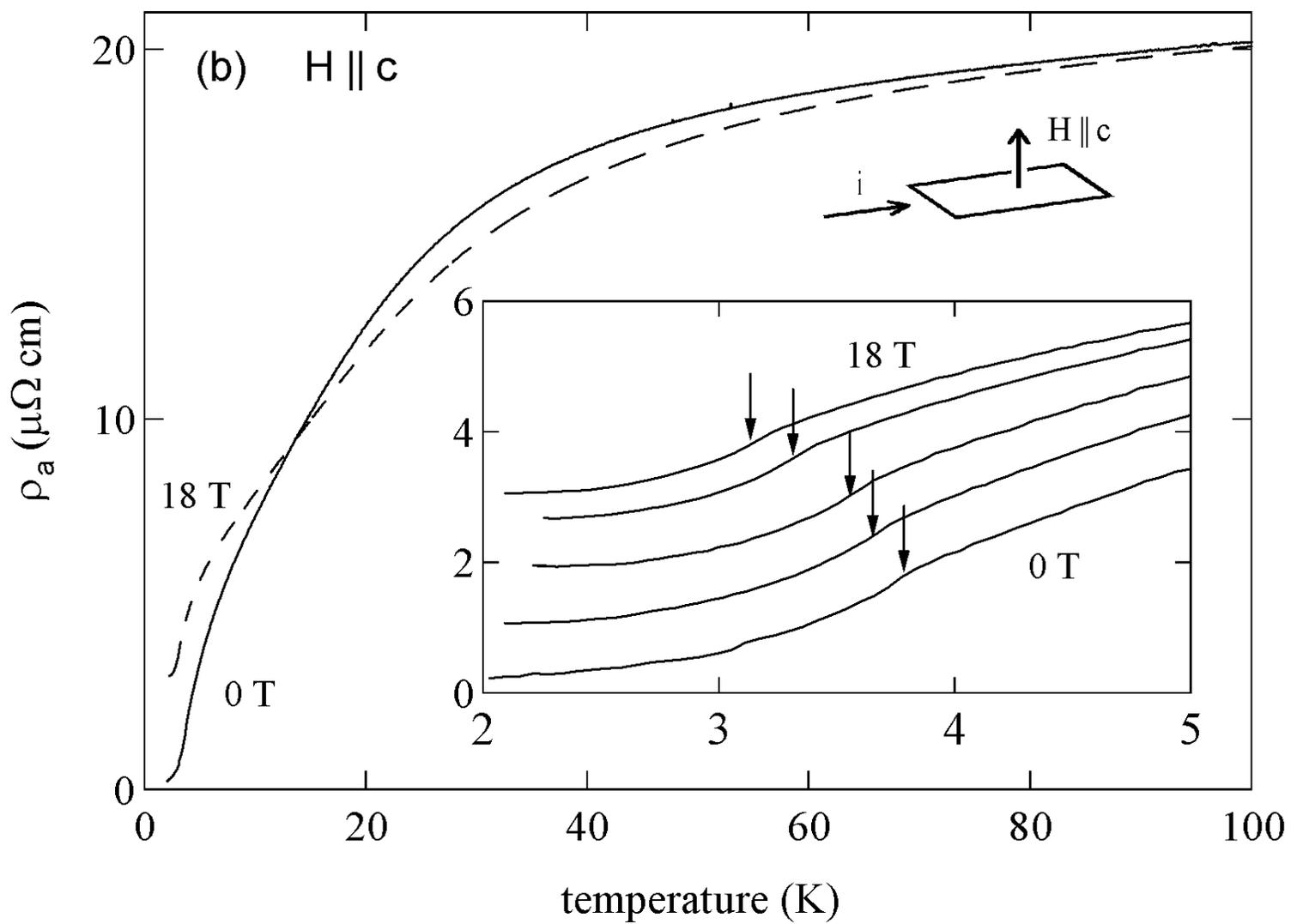

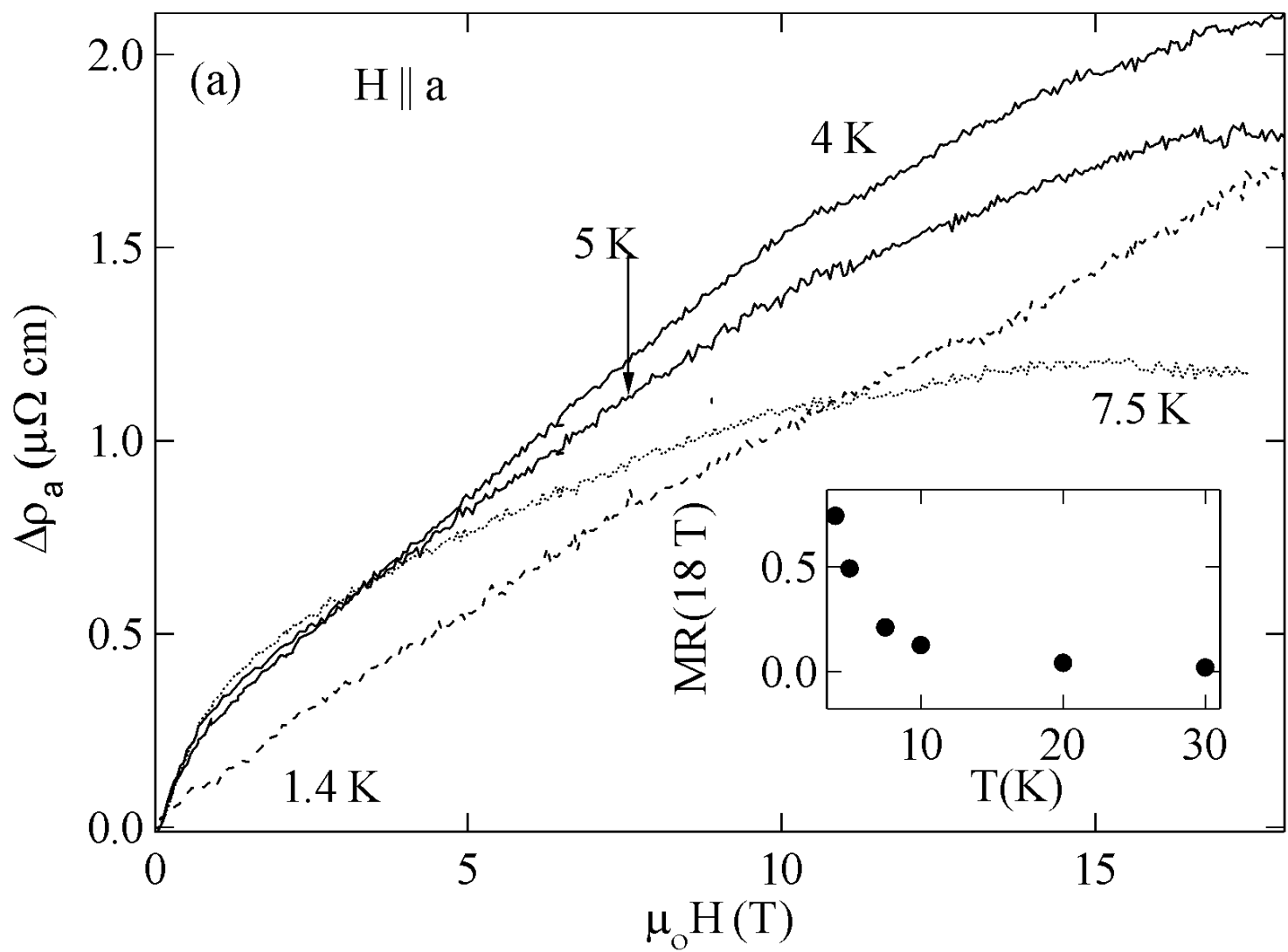

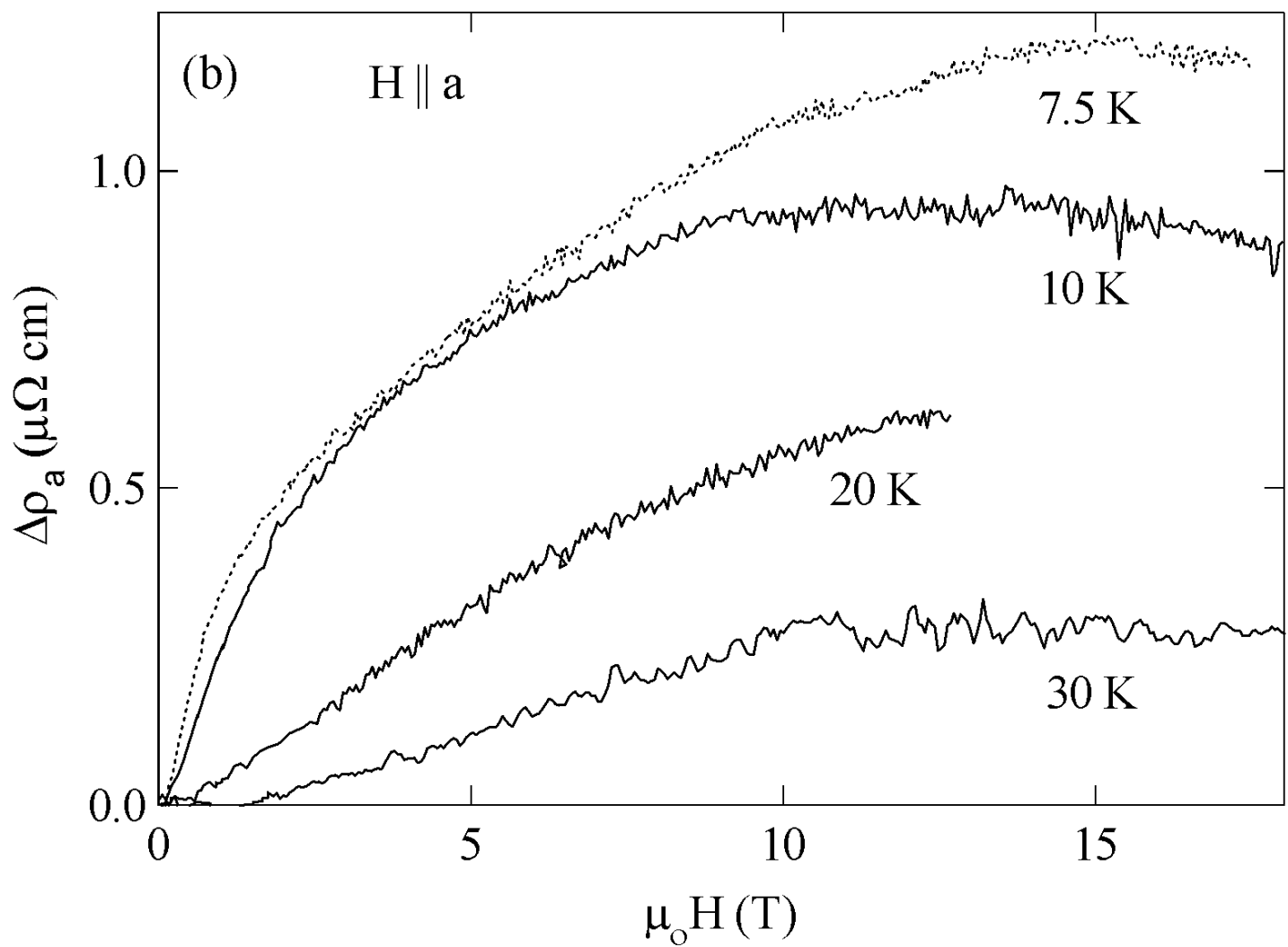

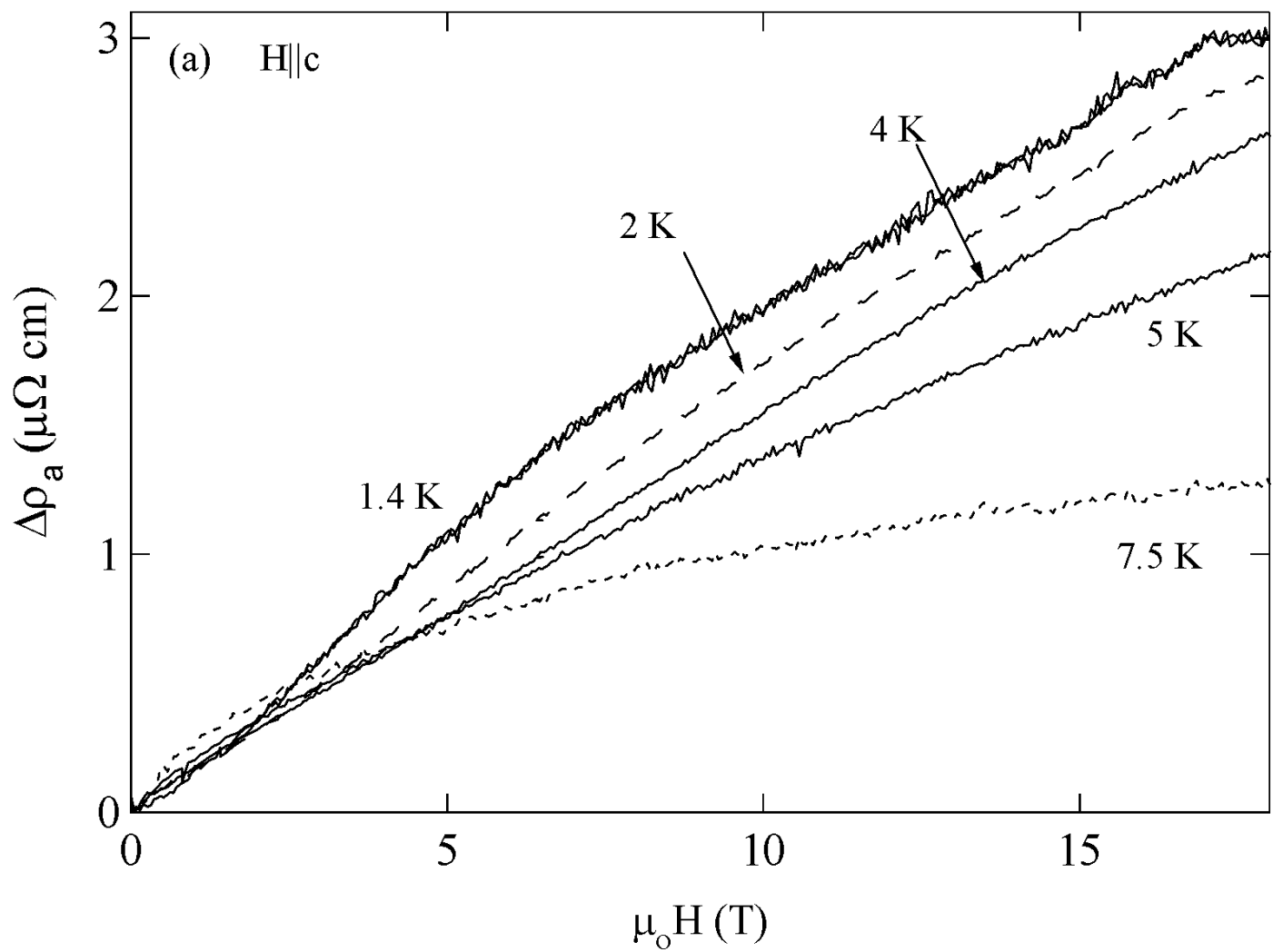

(a)  H‖c

$\Delta\rho_a$ ($\mu\Omega$ cm)

1.4 K

2 K

4 K

5 K

7.5 K

$\mu_o H$ (T)

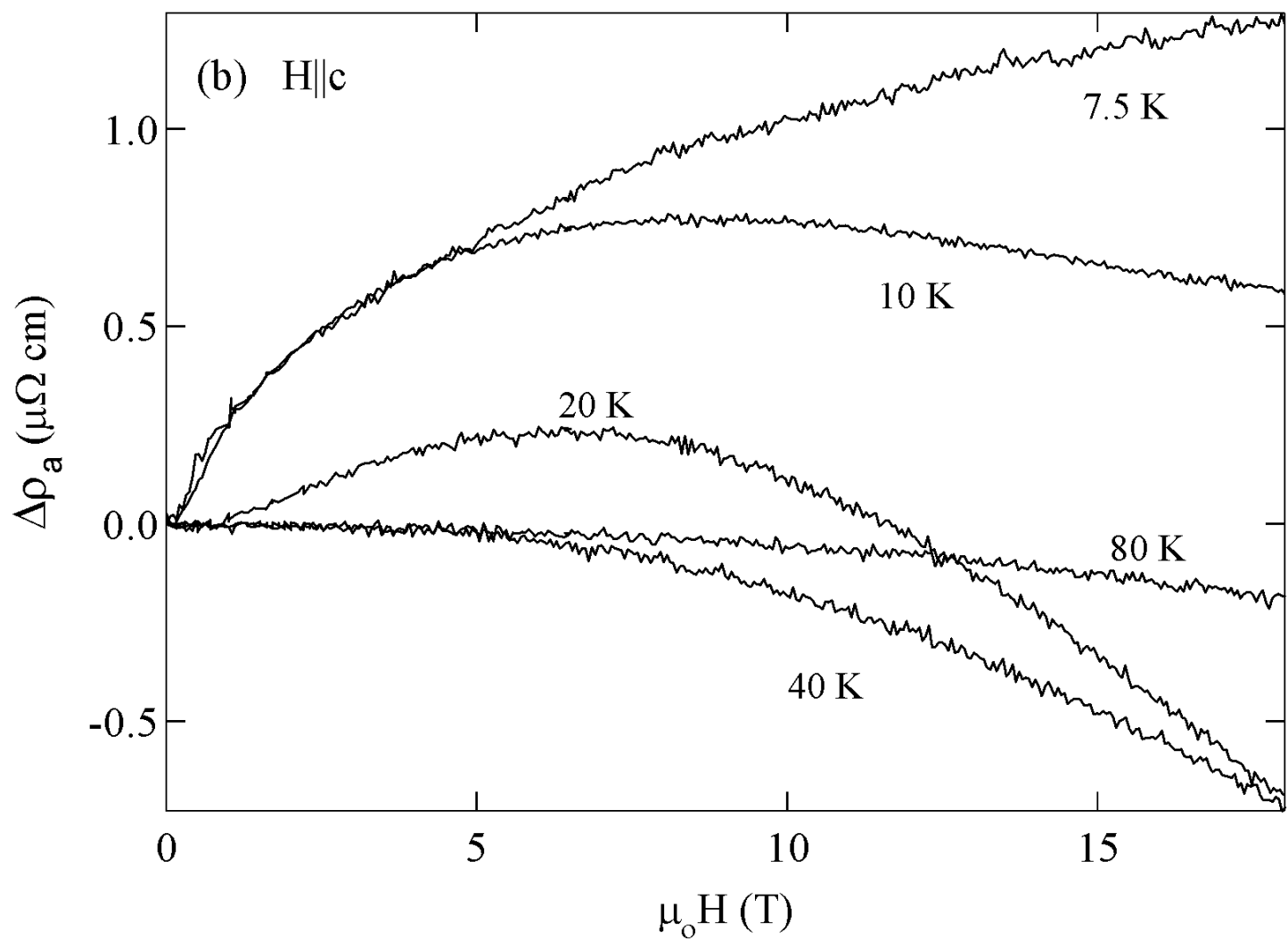

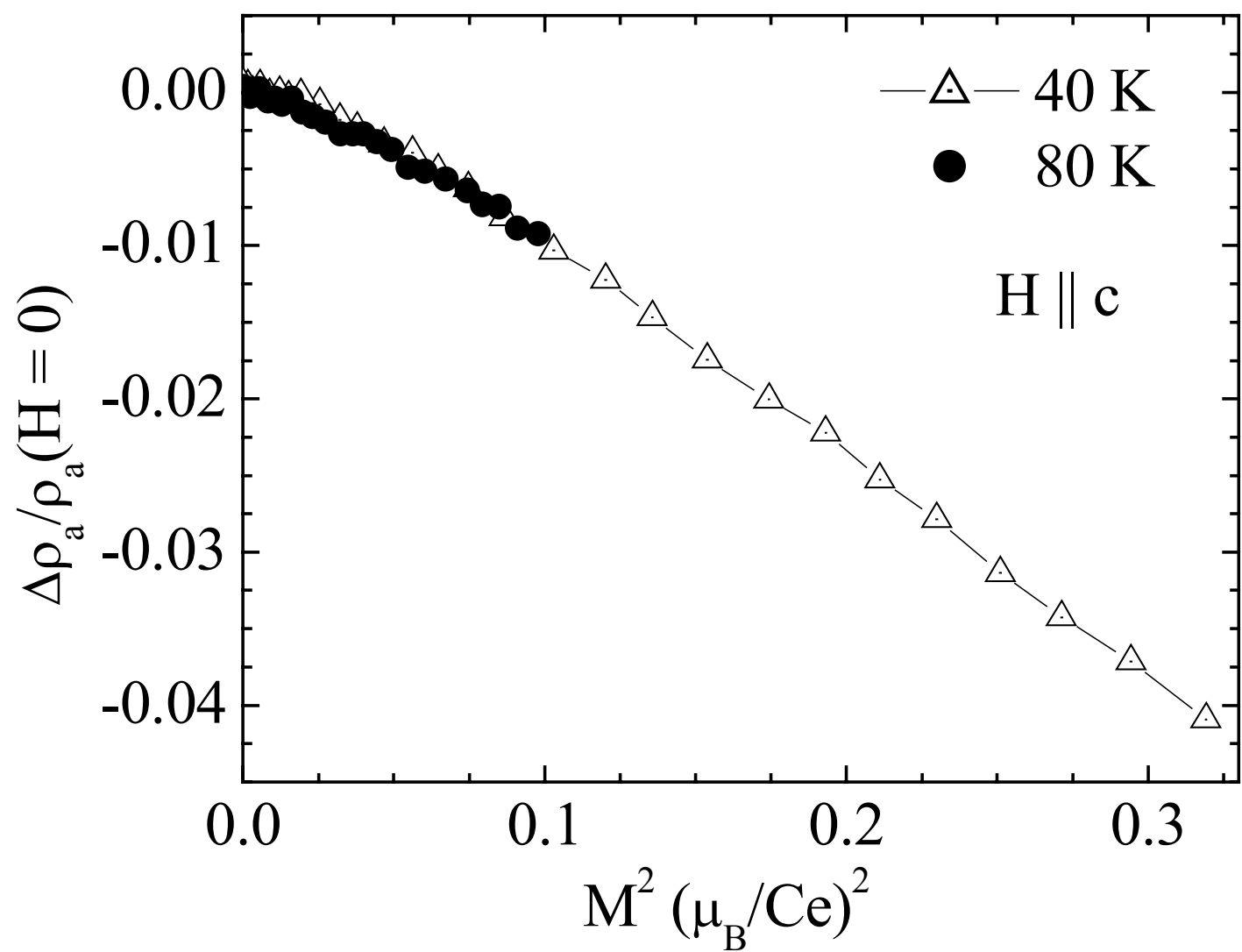